\documentclass[twocolumn,showpacs,preprintnumbers,amsmath,amssymb,unsortedaddress]{revtex4}
\usepackage{txfonts}
\usepackage{color}

\usepackage{subfigure}
\usepackage{graphicx}
\usepackage{dcolumn}
\usepackage{bm}

\begin{document}

\title{Novel phase transition in collective motion with appearance of abnormal agents}

\author{Han Yan}\email{hanyan@mail.ustc.edu.cn}
\affiliation{Department of Modern Physics, University of Science and Technology of China, Hefei 230026, China}

\author{Yu-Jian Li}\email{jinzhili@mail.ustc.edu.cn}
\affiliation{China Satellite Maritime Tracking and Control
Department Jiangyin, 214400, China}
\affiliation{Department of Modern
Physics, University of Science and Technology of China, Hefei
230026, China}

\author{Zhen-Dong Xi}
\affiliation{China Satellite Maritime Tracking and Control
Department Jiangyin, 214400, China}

\author{Bing-Hong Wang}\email{bhwang@ustc.edu.cn}
\affiliation{Department of Modern Physics, University of Science and
Technology of China, Hefei 230026, China}

\begin{abstract}
We introduce a novel type of \emph{abnormal agents} that proceed in the opposite direction of that defined for the normal agents. A new order parameter, $y$, is introduced to describe the characteristic of the system. Many interesting phenomenons emerge as the number of abnormal agents number shifts, e.g., the system may transform into a new phase (from $y\sim 1$ to $y\sim -1$) suddenly with abnormal agents getting denser, or stay disordered forever. A variety of other properties like size effect, agent speed and sight radii that  have impacts on the new collective dynamics are also studied in detail. We suggest that our model or its modified versions can be applied to explain a variety of phenomenons with multiple kinds of particles interacting with each other and shape the system dynamics.
\end{abstract}

\date{\today}


\maketitle

\section{Introduction}
\emph{Collective motion} is a fascinating phenomenon that emerges in the nature from microscopic scale like molecular motors\cite{b.0.01} , bacteria\cite{b.0.02,b.0.03,b.0.04}and individual cells\cite{b.0.05,b.0.06}, to macroscopic level such as insect swarms\cite{b.0.07} and bird flocks\cite{b.0.08}.\\

In 1995, Vicsek \emph{et al}. have proposed an elegant and powerful model to describe such behaviors \cite{b.10}. Due to its elegancy and efficiency, Vicsek Model (VM) has received a wide attention, and its mutations have been extensively discussed. In Ref.\cite{b.20}, impacts of alignment rules or inelastic collisions were studied. Ref.\cite{b.30} studied the influence of  simultaneously presented volume exclusion and self-propulsion. The collective motion of polar units moving in two dimensions with nematic collisions was presented in Ref.\cite{b.40}. In Ref.\cite{b.50,b.60,b.70,b.80} adhesion between the particles was introduced to avoid isolation of clusters under open boundary conditions. The optimal view angle for the formation of collective motion was studied in Ref.\cite{b.90,b.91}. Adaptive speed to accelerate flocking was studied in Ref.\cite{b.100}. Ref.\cite{b.110} provided a comprehensive review of the development of research in collective motion, from the basic ideas to exciting recent discoveries.\\

In most of these modified versions, there are usually only one type of self-propelled agents, i.e., all the agents share the same properties. Nevertheless, it is interesting to question that, if we place agents with different behaviors together, how will their interactions shape the system dynamics. Introducing different types of agents in a modified VM Model may provide us a potential tool kit for a wider range of research topics. For example,the formation of mutual cooperative phenomenon, the spread of infectious disease between different species, a unique form of colony shaped by the interaction between two types of coexisting bacteria and the diffusion of atoms in a certain kind of material. Therefore we propose that further study of the collective dynamics with multiple types of self-propelled particles may both deeper our understanding in theoretical nonequilibrium statical physics and provide useful methods for practical modeling of the nature.\\

This paper is arranged as follows. In Sec.\ref{secmodel}, we describe the rules of out model in detail, and illustrate the states the system may be in. In Sec.\ref{secphase}, we present how the phase transition occur with parameters running. In Sec.\ref{secfurther}, we inspect further into the system dynamics with a number of numerical simulations. At last, a summary is given.\\

\section{Model With Abnormal Agents}\label{secmodel}
\subsection{Model description}
In our model, all the agents are assumed to move in a square-shaped cell of linear size L with periodic boundary conditions.
Initially, each agent is randomly distributed in the square, with its velocity direction randomly distributed in $[\pi,-\pi)$.
The position of the ith agent is updated as
\begin{equation}\label{x}
\vec{x}_i(t+1) = \vec{x}_i(t) + \vec{v}_i(t)
\end{equation}
at each time step.
For a normal agents, its direction is defined as in VM:
\begin{equation}\label{theta}
\theta_i(t+1)=\langle \theta_i(t) \rangle_r + \Delta \theta_i,
\end{equation}
where $\langle \theta_i(t) \rangle_r$ is the direction of the average velocity of the agents within the horizon radii $r$ of the $i$th agent, including the $i$th
agent itself. It is obtained from:
\begin{equation}\label{tantheta}
 \langle \theta_i(t) \rangle_r = \arctan [\langle v_i\sin\theta_i(t) \rangle_r/\langle v_i\cos\theta_i(t) \rangle_r].
\end{equation}
In Eq.\ref{theta}, $\Delta \theta_i$ is a random number
 evenly distributed in $[-\eta,\eta)$, denoting the thermal noise of the agents.
 Unless otherwise noted, $\eta$ is set as $0$. In other words, we mainly focus on analyzing noise-free systems.\\
Besides the normal agents, we also introduce a new type of agents named the \emph{abnormal agents}. The only difference between the normal agents introduced in the VM and the abnormal agents we introduce is that, the direction of the velocity of an abnormal agent at time $t+1$ is opposite to the direction of the average of the velocity of agents around it at time $t$:
\begin{equation}\label{thetaab}
\theta_i(t+1)=-\langle \theta_i(t) \rangle_r + \Delta \theta_i.\\
\end{equation}

This new property of the abnormal agents is simple but nontrivial. The dynamic of a single abnormal agent is no more complicated than a normal one. However, this property makes the abnormal agents behave completely contrary to the normal ones in the same environment. We choose this property not only because we want to study the system with two type of particles behaving in a totally contrary way, but also because it can be a representative example for further studies involving abnormal agents with different properties.\\

In each simulation, we place $n_{ab}$ abnormal agents and $n_{nm}$ normal agents in the system. In the latter section we will demonstrate how phase transition occur with $n_{ab}$ (and $n_{nm}$, sometimes) shifting.\\
We define the order parameter $\varphi$ as the normalized average velocity:
\begin{equation}\label{order}
\varphi_{nm(ab)}=\frac{1}{N_{nm(ab)}v_0}\left |{\sum_{i\in nm(ab)}\vec{v}_i}\right |,
\end{equation}
where the subscript $nm$ and $ab$ stand for the normal agents and the abnormal agents, respectively.\\
A new parameter, $y$, is introduced in our work to depict which phase the system is in. It is defined as:
\begin{equation}\label{y}
y_{nm(ab)}=\frac{1}{N_{nm(ab)}^2v^2_0}(\sum_{i\in nm(ab)}\vec{v}_i(t))(\sum_{i\in nm(ab)}\vec{v}_i(t-1)).
\end{equation}
From the expression one will see that $y$ always satisfies $-1\leqslant y\leqslant1$.

\subsection{Different phases of the system state}
\begin{figure}
\centering
 \subfigure[Phase $1$: $n_{nm}=800$, $n_{ab}=200$]
 {
   \includegraphics[width=260 px] {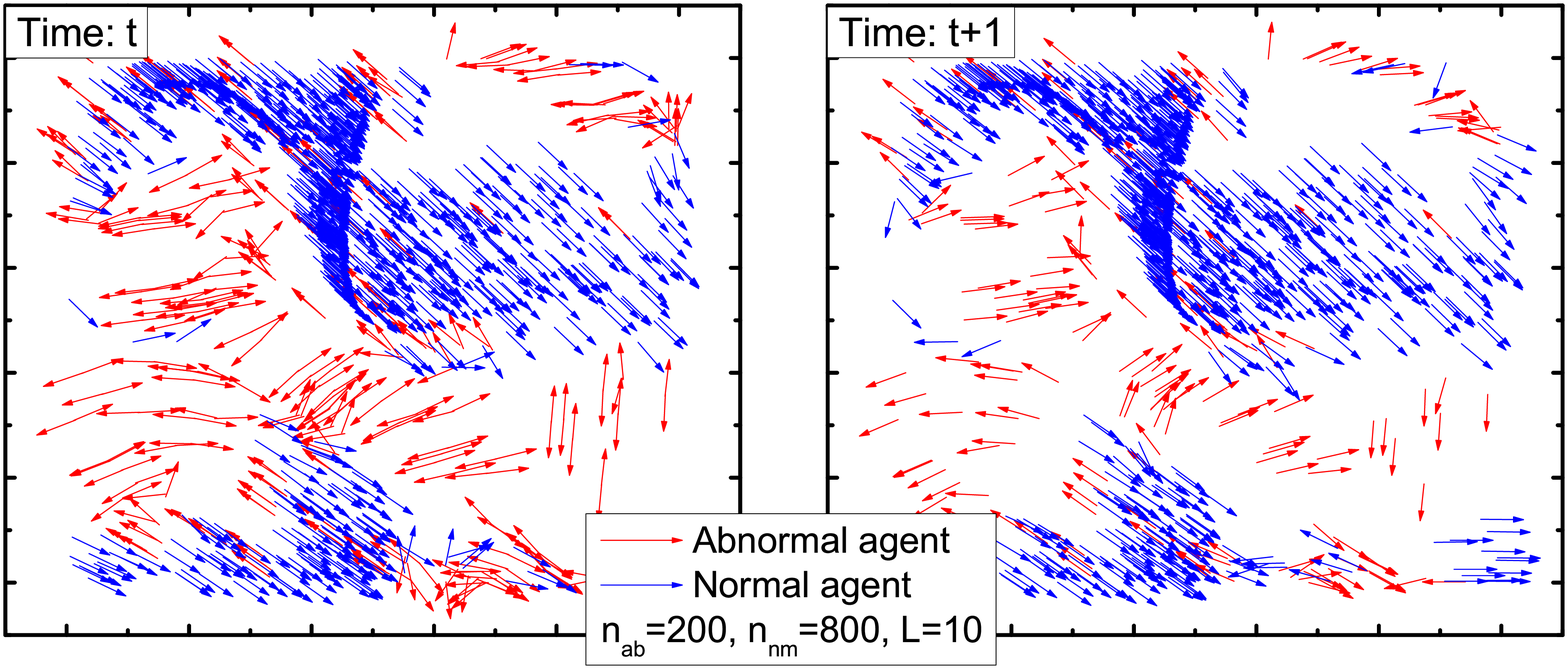}
   \label{phase:1}
}
 \subfigure[Phase $2$: $n_{nm}=200$, $n_{ab}=800$]{
   \includegraphics[width=260 px] {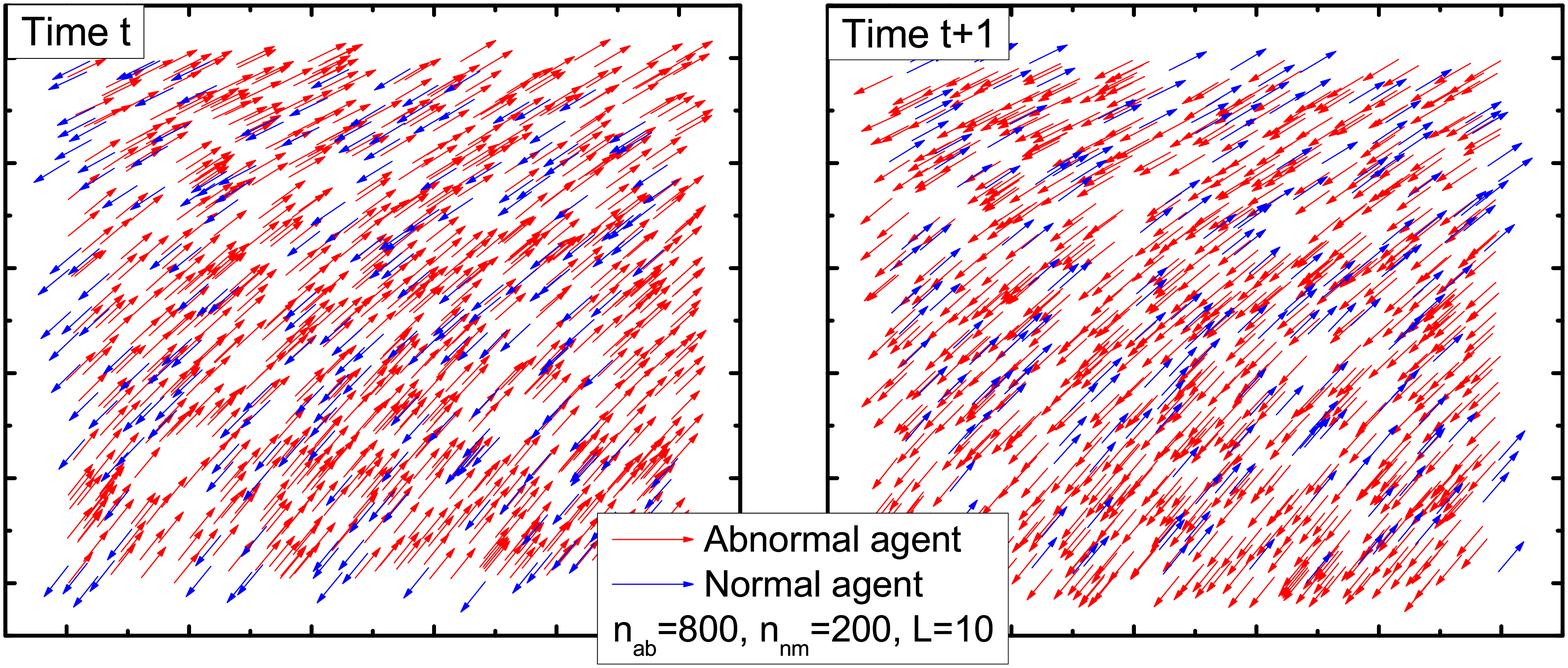}
   \label{phase:2}
 }
\caption{Phase $1$ and $2$ collective motion states in two consecutive time steps}\label{phase}
\end{figure}

With $n_{ab}$ and $n_{nm}$ shifting, we can observe the transition between two phases of collective motion. One of them is similar to the collective motion in VM, while the other is a novel type that has never been reported. The representative figures are presented in Fig.\ref{phase}\\

\begin{figure}
\centering
 \subfigure[Collective motion with $n_{nm}=570$, $n_{ab}=930$]{
   \includegraphics[width=260 px] {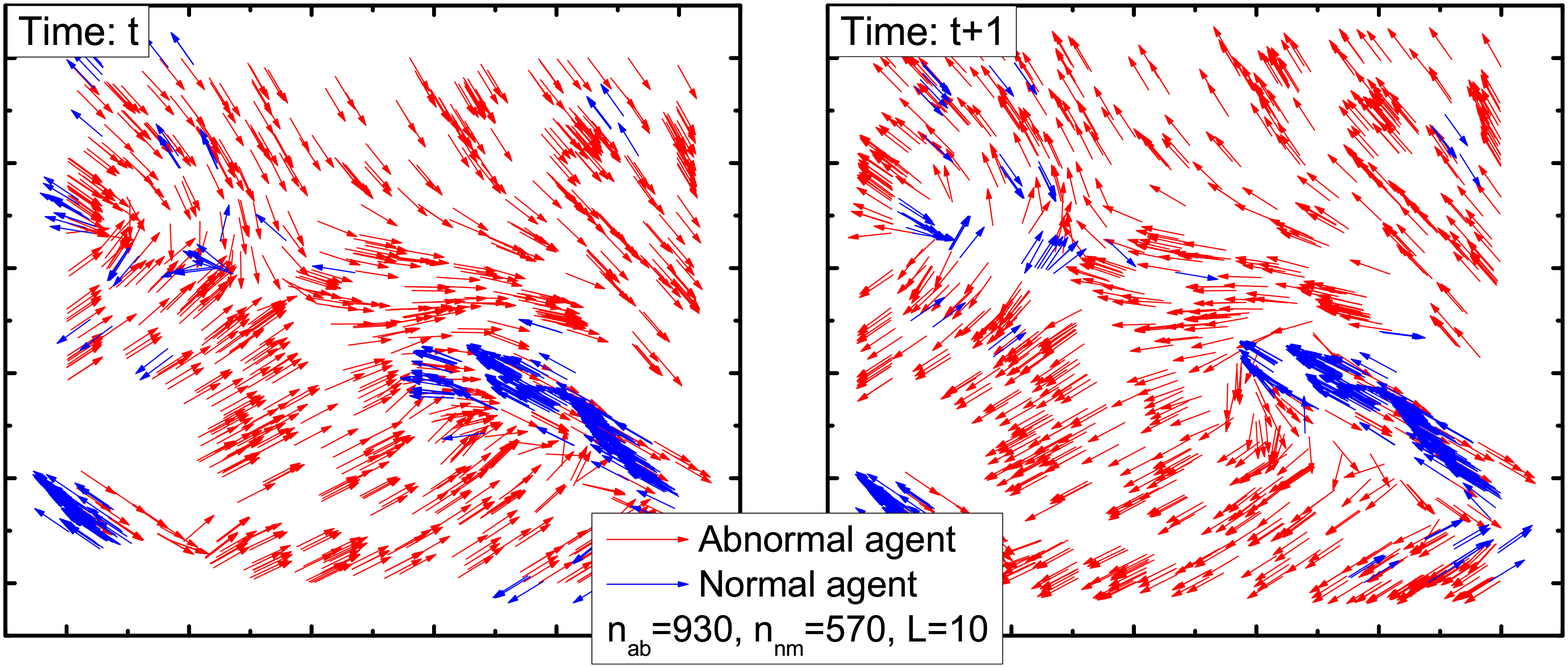}
   \label{tran:1}
 }
 \subfigure[Collective motion with $n_{nm}=530$, $n_{ab}=970$]{
   \includegraphics[width=260 px] {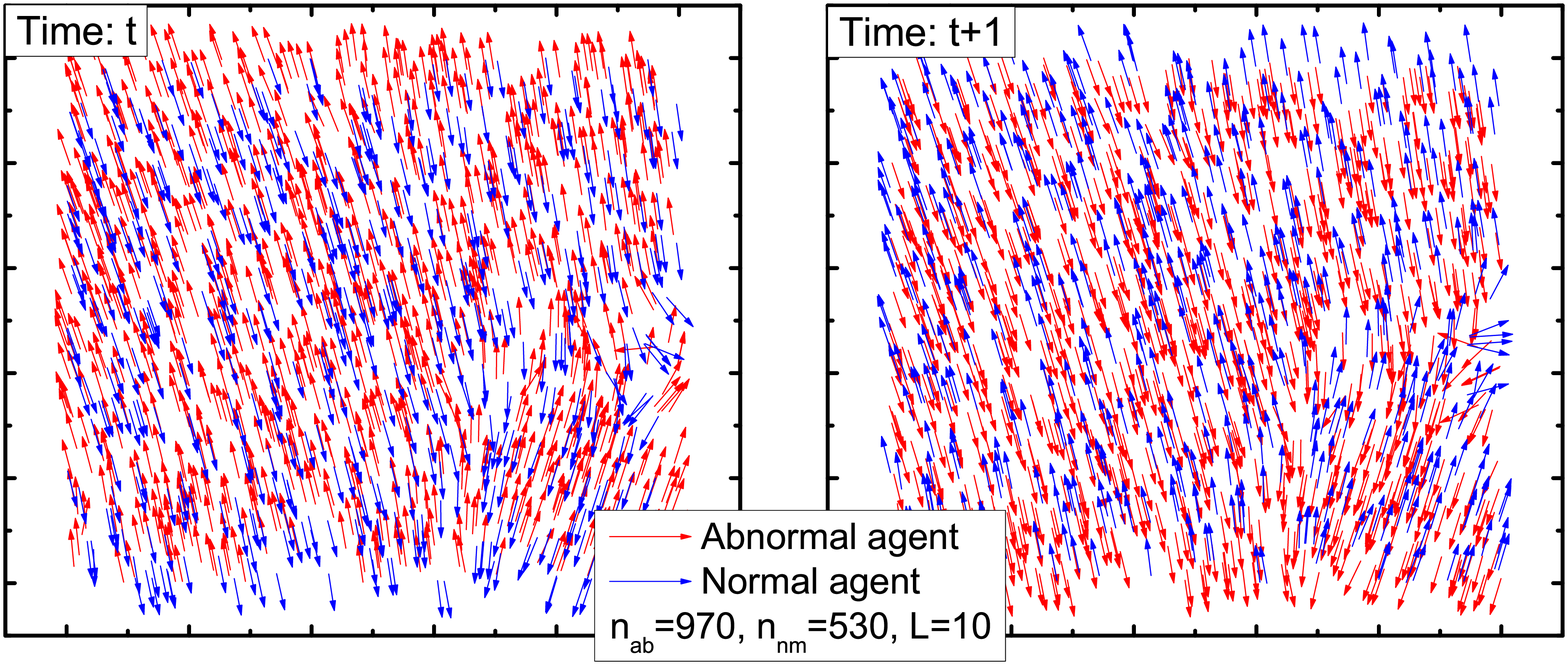}
   \label{tran:2}
 }
\caption{Collective motion states near phase transition point}\label{tran}
\end{figure}

The first phase (referred to as \emph{Phase 1}) of collective motion, similar to the one appears in the VM, will happen in our model when $n_{ab} < n_{nm}$ or $n_{ab}\backsim n_{nm}$. In this phase, all the normal agents form a steady flow toward one direction, as illustrated in Fig.\ref{phase:1} in two consecutive time steps. Obviously $y_{nm}\backsim 1$ in this phase.\\

The second phase (referred to as \emph{Phase 2}) happens when $n_{ab} > n_{nm}$ to a certain extent. This is a novel state that does not exist in literature. In phase 2 the abnormal agents outnumbered the normal ones and dominate the collective dynamics of the system. Fig.\ref{phase:2} illustrates this state in two consecutive time steps. \\
\begin{enumerate}
  \item In the first step, all the abnormal agents head toward the same direction, and the normal agents towards the opposite.\\
        In the following step, as abnormal agents will adjust their direction opposite the average direction of the previous step, which is actually the the average direction of the abnormal ones as they outnumber the normal ones, all of them turn around and head opposite to their previous direction.
  \item So do the normal agents as they tend to follow the average direction of the agents around each agent in the first step.
  \item In the third step, the system goes back to the state in step one, and this circulation goes on forever.
\end{enumerate}
   In this phase, as both types of agents are always turning around, we have $y_{nm}\backsim -1$. Therefore, $y_{nm}$ can be used as a parameter to describe the phase transition.\\

To further illustrate the system dynamics near the phase transition points, we present Fig.\ref{tran}. Fig.\ref{tran:1} illustrate a phase $1$ state near the transition point. One can see that the abnormal agents in the area free from the flow of normal agents have formed states similar to phase $2$, but locally in small groups. The normal agents form a cluster dense enough to sail through the sea of abnormal agents, which would bring the normal agents into phase $2$ if the normal agents is not dense enough. This happens just after we increase $n_{ab}$ by $50$ and decrease $n_{nm}$ by 50. The system transformed into phase $2$ is depicted in Fig.\ref{tran:2} .\\

\begin{figure}
  \includegraphics[width=260 px]{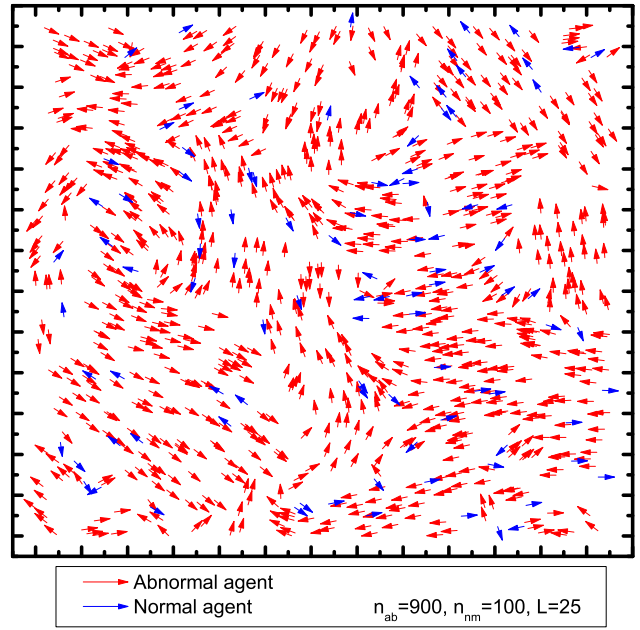}\\
  \caption{Disordered state, with $n_{nm}=100$, $n_{ab}=900$ and $L=25$}\label{l25distribution}
\end{figure}

Besides forming collective motion, the system can also stay disordered forever, if the densities of the abnormal agents $\rho_{ab}$ and the normal agents $\rho_{nm}$ are low enough, as illustrated in Fig.\ref{l25distribution}. In this situation, the abnormal agents dominate the system dynamic, and form phase $2$-like state locally. However, the information of the direction of velocity cannot spread to the globe, and the whole picture is still of no order.\\

\section{Phase transitions}\label{secphase}
\subsection{Phase transition with respect to $n_{ab}$}
\begin{figure}
  \includegraphics[width=260 px]{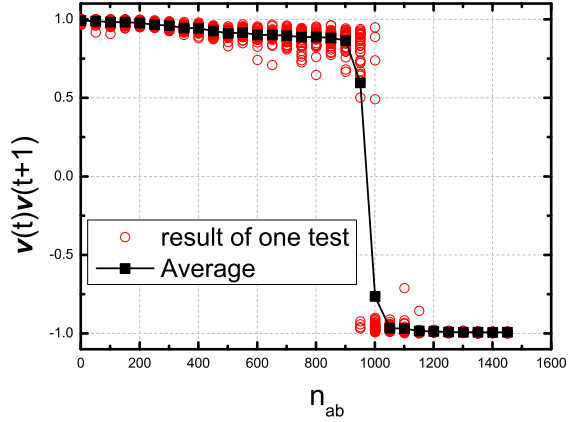}\\
  \caption{First order phase transition of $y$ with growing $n_{ab}$}\label{yorder}
\end{figure}

As indicated in Fig.\ref{phase} and \ref{tran}, the phase transition appears after $n_{ab}$ outnumbers $n_{nm}$. In the previous section we have proposed $y_{nm}$ as a parameter to describe the phase the system is in. Fig.\ref{yorder} illustrates the relation between $y_{nm}$ and $n_{ab}$, with $n_{nm}$ fixed as $500$. This just reflect the fact that the ratio of $n_{ab}$ and $n_{nm}$ decides which type of agents paly the dominant role of the collective dynamic, so the collective dynamic shows different patterns with respect to changing agents number ratio.\\

\subsection{Phase transitions with respect to size effect}

Size effect is an important topic when discussing self-propelled particles. We are aware whether the system dynamic is directly related to $L$ or not, and will demonstrate that this relation do not exist in the following.\\

\begin{figure}
  \includegraphics[width=260 px]{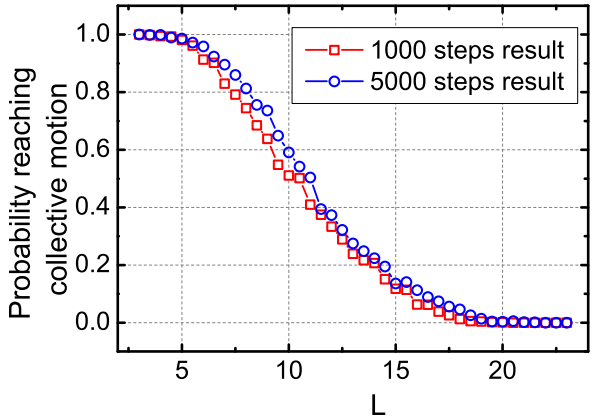}\\
  \caption{Times reach collective motion with respect to $L$. At each value of $L$ 500 ?? runs have been performed}\label{lreach}
\end{figure}

In Fig.\ref{l25distribution}, one may notice that not only the ratio, but also the absolute values of $\rho_{ab}$ and $\rho_{nm}$ play a critical role in the formation of collective motion. For phase $1$, we have already understood its similarity with the VM. So in this subsection we only focus on the discussion of the situation in phase $2$.\\
 In Fig.\ref{lreach} , we fix $n_{ab}=950$ and $n_{nm}=50$, and let $L$ run from $3$ to $23$.
One can find that with $L$ increasing and density of agents decreasing, the system is less likely to form collective motion. This reflects the fact that when $\rho_{ab}/\rho_{nm}$ is large but $\rho_{ab}$ too small, the phase $2$ state cannot form globally. Fig.\ref{l25distribution} gives one typical screenshot of this state.\\

\section{Further inspection into the system dynamics}\label{secfurther}
\noindent\textbf{The influence of $L$}\\
\begin{figure}
  \includegraphics[width=260 px]{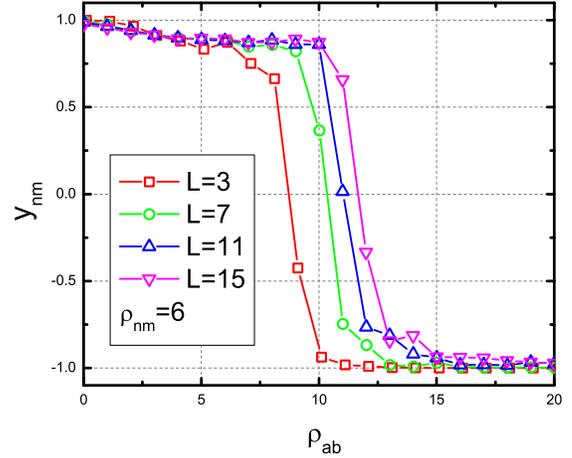}\\
  \caption{$y$ with respect to $\rho_{ab}$, with different values of $L$.}\label{xdenyproll}
\end{figure}
Fig.\ref{xdenyproll} illustrates the phase transition with respect to $\rho_{ab}$, with different values of $L$. In the simulations, $\rho_{nm}$ is fixed as 6. One can find that with $L$ running larger to infinity, the curve become stable. This proves the phase of the system to be determined by the ratio of $\rho_{nm}$ and $\rho_{ab}$. (Of course, according to Fig.\ref{lreach}, the absolute value of $\rho_{nm}$ and $\rho_{ab}$ are also critical.) Further more, the system dynamics is not directly related to $L$, which leads to a nontrivial conclusion that the phenomenon we describe can emerge at any scale.\\

\noindent\textbf{The influence of velocity}\\
\begin{figure}
  \includegraphics[width=260 px]{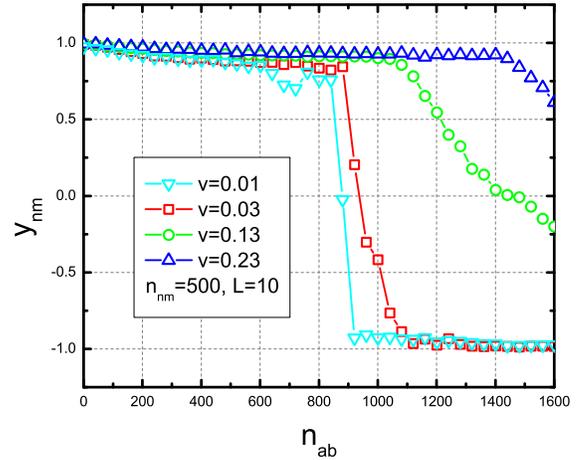}\\
  \caption{$y$ with respect to $n_{ab}$, with different values of $v$.}\label{xabyprolvel}
\end{figure}
Besides L, we are also interested to see how the value of $v$ can influence the system dynamics.
Fig.\ref{xabyprolvel} illustrates how the critical points moves to the right side with $v$ increasing. What's more, the range in which the system has a chance to stay in either phase $1$ and phase $2$ gets larger, e.g., when $v=0.01$, there is only one point with $-0.7<y<0.7$, but there are three when $v=0.03$. Each of these points represents average of 50 runs, and each run produces a $y$ with $|y|\sim 1$. This implies that with $v$ vanishing, the phase transition goes more sharply.\\

\noindent\textbf{The Influence of sight radii}\\
\begin{figure}
  \includegraphics[width=260 px]{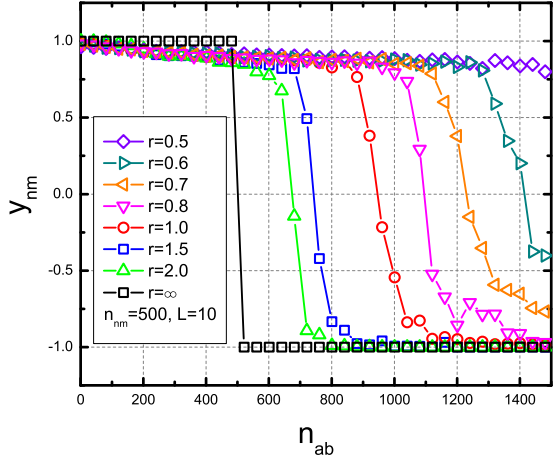}\\
  \caption{$y$ with respect to $n_{ab}$, with different values of $r$.}\label{xabyprolsight}
\end{figure}
Fig.\ref{xabyprolsight} depicts how the critical point change with respect to the sight radii $r$. With a larger horizon, the phase transition emerges earlier when $n_{ab}$ increases. The curve also goes down more sharply with a larger $r$. With $r$ running to the infinity, the phase transition became of first order, and happen exactly when abnormal agents outnumber normal agents. When $r=\inf$, the phase transition happens exactly at $\rho_{ab}=\rho{nm}$. This is understandable, since all the agents share the same $\langle \theta_i(t) \rangle$ when $r=\inf$, and $\langle \theta_i(t) \rangle$ is completely decided by the side with more agents. \\

\noindent\textbf{The influence of thermal noise}\\
\begin{figure}
  \includegraphics[width=260 px]{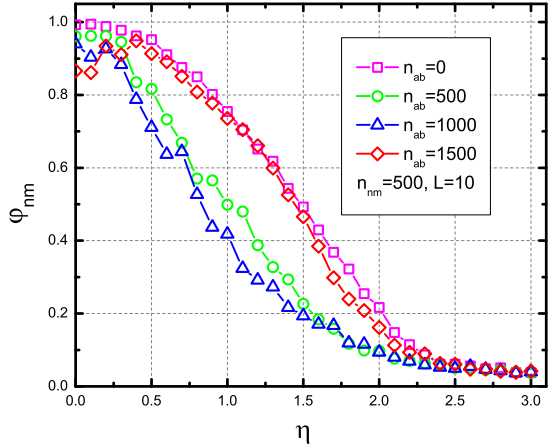}\\
  \caption{$\varphi_{nm}$ with respect to $\eta$, with different values of $n_{ab}$.}\label{xperyorderlabnm}
\end{figure}
In the previous analysis, the thermal noise $\eta$ is not taken into consideration. At the end of this paper, we present Fig.\ref{xperyorderlabnm} to show the impact on $\varphi_{nm}$ from $\eta$.
One can conclude from Fig.\ref{xperyorderlabnm} that when $n_{nm}$ and $n_{ab}$ bring the system near the critical point, e.g., $n_{ab}=500$ or $n_{ab}=1000$, the system become more likely to be disordered with $\eta$ increasing. On the contrary, when $n_{ab}=0$ or $n_{ab}=1500$, the system stays far from the critical points and shows in a higher capability to remain stable.

\section{Summary}
We have introduced a novel type of agents in the self-propelled particle system. We have described the rules of the model in detail, and introduced the concept of abnormal agents and order parameter $y$.
With $y$ indicating the characteristic of the system, we have discovered some new types of phase transitions with respect to the values of $n_{nm}$ and $n_{ab}$. The abnormal agents play the role of blocking the normal agents and impeding the formation of the previously known collective motion. However they can also promote the formation of another type of collective motion with their number growing. What's more, we have studied the properties of the system dynamics in detail, and presented a variety of numerical simulation results. We suggest that the model we introduce may imply a new research sub-direction: the interaction of agents with different behaviors in one system.\\

\acknowledgments We would like to give our appreciations to Wen-Yao
Zhang and Yuki Nagato for their kindly help. This work is funded by
the National Natural Science Foundation of China (Grant Nos.
9102402610975126, 10635040 and 11005001).

\end{document}